\pdfoutput=1
\documentclass[fleqn,usenatbib,usedcolumn]{mnras}
\usepackage{graphicx,amssymb,cite,amsmath}
\usepackage[authoryear]{natbib}
\usepackage{txfonts}
\usepackage{graphics,times,placeins,float}
\usepackage{pdflscape}
\usepackage{longtable,lscape,rotating}
\usepackage{supertabular,caption,multicol}
\usepackage{threeparttable}
\usepackage[T1]{fontenc}
\usepackage{ae,aecompl}

\def\swift{\textit{Swift}}

\def\cxo{\textit{Chandra}}
\def\hst{\textit{HST}}
\def\vtc{VTC\,J095517.5+690813}

\title{Discovery of a radio transient in M81}
\author[G. E. Anderson et al.]{G. E. Anderson,$^{1}$\thanks{E-mail: gemma.anderson@curtin.edu.au}
J. C. A. Miller-Jones,$^1$
M. J. Middleton,$^2$
R. Soria,$^{3,1}$
\newauthor
D. A. Swartz,$^4$
R. Urquhart,$^1$
N. Hurley-Walker,$^1$
P. J. Hancock,$^1$
R. P. Fender,$^5$
\newauthor
P. Gandhi,$^2$
S. Markoff,$^{6}$
T. P. Roberts$^{7}$
\\
$^1$International Centre for Radio Astronomy Research, Curtin University, GPO Box U1987, Perth, WA 6845, Australia\\
$^2$Department of Physics and Astronomy, University of Southampton, Highfield, Southampton SO17 1BJ, UK\\
$^3$College of Astronomy and Space Sciences, University of the Chinese Academy of Sciences, Beijing 100049, China\\
$^4$Astrophysics Office, NASA Marshall Space Flight Center, ZP12, Huntsville, AL 35812, USA\\
$^5$Astrophysics, Department of Physics, University of Oxford, Keble Road, Oxford OX1 3RH, UK\\
$^6$Anton Pannekoek Institute for Astronomy/GRAPPA, University of Amsterdam, 1098 XH Amsterdam, the Netherlands\\
$^{7}$Centre for Extragalactic Astronomy, Durham University, Department of Physics, South Road, Durham DH1 3LE, UK\\
}

\begin{document}

\date{Accepted XXX. Received YYY; in original form ZZZ}

\pagerange{\pageref{firstpage}--\pageref{lastpage}} \pubyear{2019}

\maketitle

\label{firstpage}

\begin{abstract}

We report the discovery of a radio transient in the spiral galaxy M81. The transient was detected in early 2015 as part of a two year survey of M81 made up of 12 epochs using the Karl G. Jansky Very Large Array. While undetected on 2014 September 12, the source was first detected on 2015 January 2, from which point it remained visible at an approximately constant luminosity of $L_{R,\nu} = 1.5 \pm 0.1 \times 10^{24}$\,erg\,s$^{-1}$\,Hz$^{-1}$ at the observing frequency of 6\,GHz for at least 2 months. 
Assuming this is a synchrotron event with a rise-time between 2.6 and 112\,days, the peak luminosity (at equipartition) corresponds to a minimum energy of $10^{44} \lesssim E_{\mathrm{min}} \lesssim 10^{46}$\,erg and jet power of $P_{\mathrm{min}} \sim10^{39}$\,erg\,s$^{-1}$, which are higher than most known X-ray binaries. 
Given its longevity, lack of short-term radio variability, and the absence of any multi-wavelength counterpart (X-ray luminosity $L_{x} \lesssim 10^{36}$\,erg\,s$^{-1}$), it does not behave like known Galactic or extragalactic X-ray binaries. 
The M81 transient radio properties more closely resemble the unidentified radio transient 43.78+59.3 discovered in M82, which has been suggested to be a radio nebula associated with an accreting source similar to SS 433. One possibility is that both the new M81 transient and the M82 transient may be the birth of a short-lived radio bubble associated with a discrete accretion event similar to those observed from the ULX Holmberg II X-1. However, it is not possible to rule out other identifications including long-term supernova shockwave interactions with the surrounding medium from a faint supernova or a background active galaxy. 

\end{abstract}


\section{Introduction}

As we approach the era of the Square Kilometre Array (SKA), astronomers are investigating how to use the low- and mid-frequency facilities for transient radio astronomy. 
Such radio observations allow us to directly probe particle acceleration and shock physics from explosive and outbursting astronomical sources, which provides direct insight into the velocity of the ejecta, the associated magnetic fields, the total energy budget, and the density distribution of the circumstellar and intergalactic media \citep{fender15ska}. The majority of radio transient studies have focused on blind searches for transients in both dedicated radio surveys and archival data \citep[e.g.][]{bower07,bower11,bannister11,ofek11,frail12,jaeger12,bell11,bell15,stewart16,carbone16,mooley16,murphy17,bhandari18}. While important steps have been made towards optimising and automating such searches, these studies have demonstrated that at current radio telescope sensitivities, radio transients are rare and faint. However, recent activities have demonstrated alternative methods for radio transient discovery that can be used to supplement the blind survey approach, and are predicated on performing radio observations of sky regions in which we expect to find radio transients.

For this paper, we will discuss the ``targeted" (or monitoring) observational technique, which involves performing targeted radio monitoring of sky regions with a high source density such as face-on galaxies, globular clusters, the Galactic Centre or portions of the Galactic Plane, providing many potential transient sources within a single field-of-view. The Karl G. Jansky Very Large Array (VLA) observations of nearby (within a few Mpc) galaxies are sensitive to the radio emission from binary neutron star mergers detected by the Advanced Laser Interferometer Gravitational Wave Observatory (aLIGO) and the Virgo gravitiational wave interferometer \citep{mooley18a}, radio-bright core-collapse supernovae (SNe), and transient radio jets from ultraluminous X-ray sources \citep[ULXs; e.g.][]{middleton13}, which are XRBs accreting at or above the Eddington rate.

An excellent example of the targeting technique is demonstrated by the discovery of three unidentified radio transients in the starburst galaxy M82 \citep[star formation rate (SFR) of around $3.6 M_{\odot}$yr$^{-1}$; see][and references therein]{grimm03}, which were discovered via frequent and ongoing radio monitoring that was running from the 1980s until 2010 \citep{kronberg85,muxlow94,mcdonald02,fenech08,muxlow10,gendre13}. These transients were detected between 1981 and 2009, indicating a minimum occurrence rate of 1 every 10 years \citep{muxlow10}. One of these transients, 43.78+59.3 \citep[as named by][]{muxlow10}, switched on and rose to its maximum radio luminosity within 8 days, increasing in flux by a factor of 5, where it remained stable in flux for at least 19 months \citep{gendre13}. Several classifications were suggested by \citet{muxlow10}, including a faint SN or a second SMBH from a galaxy that merged with M82 sometime in its past. However, the most likely classification is a stellar mass accreting system \citep{muxlow10,joseph11}. M82 is an unusual galaxy due to its recent interaction history so may have an atypical transient event rate \citep[for example, a study of the compact radio sources within the central kpc of M82 identified 30 SN remnants and 16 H{\sc II} regions;][]{mcdonald02}. For comparison, a search for radio transients was also conducted in the spiral galaxy M51, which was made possible by the frequent radio monitoring of SN 1994I with the pre-upgrade VLA \citep[the VLA was commissioned in 1980 and upgraded in 2012;][]{perley11}. No radio transients were detected in 31 epochs spanning 5 months, probing day to monthly timescales, down to a flux density of 0.5 mJy. This resulted in a $2\sigma$ upper limit on the areal transient density of $<17$\,deg$^{-2}$ above a radio luminosity limit of $L_{R,\nu} \gtrsim 4 \times 10^{25}$ erg\,s$^{-1}$\,Hz$^{-1}$ at the distance of M51 \citep{alexander15}. This suggests that more sensitive radio observations (like those now possible with the upgraded VLA) are required to search for radio transients in nearby and more standard spiral galaxies. 

It was, however, the detection of bright radio flaring associated with the X-ray outburst of a transient XRB (microquasar) in M31 \citep{middleton13} that inspired a dedicated monitoring program of nearby galaxies with the VLA. The aim of this program was to search for similar accreting systems, which are more powerful than most of those we have studied in the Milky Way. The M31 transient exhibited long-lived radio flaring behaviour for at least $\sim30$\,d, reaching a peak radio luminosity of $L_{R,\nu} = 3 \times 10^{23}$\,erg\,s$^{-1}$\,Hz$^{-1}$ ($\nu L_{\nu} = 2 \times 10^{33}$\,erg\,s$^{-1}$ at 7.45\,GHz), placing it amongst the most radio-luminous of known transient XRBs \citep{middleton13}. Radio variability was observed on 10 min to one-day timescales, corresponding to a minimum brightness temperature of $T_{\mathrm{B}} \sim 7 \times 10^{10}$\,K, which exceeds that of the brightest Galactic black hole XRBs. For a period of $\sim40$\,d, the M31 transient was accreting in the super-Eddington regime as its X-ray luminosity was brighter than the theoretical Eddington luminosity limit for a 10\,$M_{\odot}$ black hole \citep[X-ray luminosities of $L_{X} \gtrsim 10^{39}$\,erg\,s$^{-1}$;][]{feng11}, making it a transient ultraluminous X-ray source (ULX). 
The observed radio behaviour during this period was very different to what is observed from sub-Eddington Galactic black hole XRBs, where the radio jets are quenched during the soft X-ray spectral state at the peak of the outburst \citep{fender04}. This may indicate that in the super-Eddington regime, accreting objects have a different coupling between the inflow of material and the jetted outflows than in the sub-Eddington regime. This is expected as in the super-Eddington regime it is predicted that the accretion disk gets puffed up by radiation pressure, with excess material being removed in the form of a massive and optically thick, fast-moving radiation-pressure-driven wind \citep{shakura73,abramowicz88}.
Additionally, magnetohydrodynamical simulations predict the existence of powerful, collimated jets from super-critically accreting black holes \citep{ohsuga11,mckinney14,narayan17} and neutron stars \citep{parfrey17}.

The study of ULXs in nearby galaxies is of particular interest as few such sources reach similar X-ray luminosities in our own Milky Way, and only during outburst \citep[e.g. V404 Cyg;][]{motta17}. Many ULXs are persistent X-ray sources, and while partially self-absorbed compact radio jets have never been directly detected from their core, diffuse radio emission is observed around a small number of ULXs as a result of their disk winds and/or (undetected) jets interacting with the surrounding environment. These outflows give rise to radio synchrotron nebulae or ``bubbles". Such radio bubbles have luminosities ranging from $\nu L_{\nu} \sim 2 \times 10^{34}$ to $2 \times 10^{35}$\,erg\,s$^{-1}$ at 5\,GHz and have characteristic lifetimes of $\sim10^{4}$ to $10^{5}$\,y \citep[e.g. MF16 in NGC 6946 and M51 ULX1;][]{vandyk94,urquhart18}. However, there is at least one ULX, Ho II X-1 \citep{miller05,cseh14,cseh15HoII}, that produces radio lobes with transient characteristics indicative of multiple and recent (within a few years) ejection events.   

A larger sample of transient ULXs is required to understand accretion-jet coupling at the highest accretion rates, and the impact of the jetted outflows' feedback on their surrounding environments.  
We therefore embarked on a radio monitoring campaign of the nearby spiral galaxy M81 to search for transients. 
M81 is one of largest nearby galaxies observable in the Northern Hemisphere, lying at a distance of $3.6\pm0.2$ Mpc \citep{gerke11} in an interactive group that includes M82, with a SFR between $0.3$ and $0.9 M_{\odot}$yr$^{-1}$ \citep[depending on the SFR indicator;][]{gordon04}. 
Its large size, and therefore high stellar mass predicts a high number of low mass X-ray binaries \citep{grimm02}, while being nearly face-on ensures that we are not viewing through a large absorbing column. These properties, along with its close proximity makes it an ideal galaxy to target for this monitoring survey.

Here we present the discovery of a radio transient in M81. The program made use of VLA filler time, collecting data in multiple telescope configurations at 6\,GHz, and obtaining 48 epochs between 2013 January and 2015 March. In Section 2, the VLA observations and analysis are described. Section 3 explores the radio properties of the M81 transient, including its position, luminosity, spectral index, variability, minimum energy, jet power, and magnetic field strength (assuming equipartition), along with implications for the radio transient surface density of M81. This is then followed by multi-wavelength and archival investigations at the position of the transient, particularly using new and archival \cxo\ \textit{X-ray Observatory} observations, and archival \swift\ \textit{X-ray Telescope} (XRT) and \textit{Hubble Space Telescope} (\hst) data. Transient databases and catalogues are also searched for potential counterparts and associations. Section 4 explores the possible nature of the M81 transient via comparisons with known radio transient classes and other extragalactic radio transients.

\section{Very Large Array radio observations}

\begin{figure*}
\begin{center}
\includegraphics[width=0.95\textwidth]{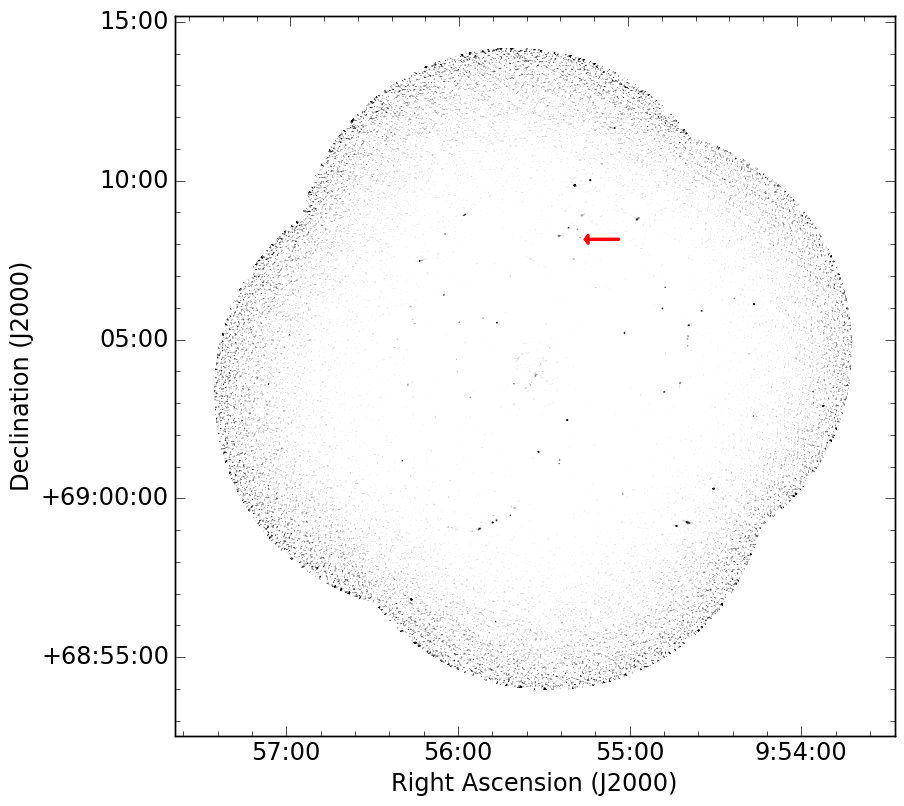}
\end{center}
\caption{Mosaicked image of the four B configuration observations (one for each quadrant) of M81 obtained during 2015 March. The transient, \vtc{}, discussed in Section~\ref{results} is indicated by the red arrow. See Figure~\ref{fig:vla} for a zoomed view of the transient. The AGN is located at the centre of the mosaic but was peeled out of the visibilities to minimise imaging artefacts.}
\label{fig:m81_vla}
\end{figure*}

The VLA monitoring campaign of M81 took place between 2013 January and 2015 March. Making use of VLA filler time, at least two epochs of data, one hour in duration, were obtained in each of the A, B, C, D and DnC configurations in the C band ($4.5-6.5$\,GHz) during this 2 year period (proposal codes 13A-256, 13B-138, 14A-167, 14B-042, and 15A-060; PI M. Middleton). In order to monitor the full galactic disk of M81, four pointings were performed (one for each quadrant) within 1 to 23 days of each other for a given configuration. The resulting mosaic of a set of four pointings covered $\sim0.08$deg$^{2}$ and was approximately centered on the central Active Galactic Nucleus (AGN) of M81 (e.g. Figure~\ref{fig:m81_vla}). For this paper, we concentrated on observations of quadrant 1 (pointing centre at $\alpha \mathrm{(J2000.0)} =  09^{\mathrm{h}}55^{\mathrm{m}}40\overset{\mathrm{s}}{.}71$ and $\delta (\mathrm{J2000.0}) = +69^{\circ}07'19\overset{''}{.}45$) and quadrant 2 (pointing centre at $\alpha \mathrm{(J2000.0)} =  09^{\mathrm{h}}54^{\mathrm{m}}58\overset{\mathrm{s}}{.}45$ and $\delta (\mathrm{J2000.0}) = +69^{\circ}04'43\overset{''}{.}90$). See \citet{wrobel16} for a full description of the quadrant pointings and observational details.

The flagging, calibration, and imaging of each observation, along with source position and flux density measurements, were conducted using the Common Astronomy Software Application package \citep[{\sc CASA};][]{mcmullin07} using standard techniques. The dominant radio source in M81 is its central AGN, with a flux density varying by a factor of $\sim2$ over the 2 years of observations. As the AGN is between $3\overset{'}{.}1$ and $3\overset{'}{.}5$ off-centre (at $\sim60-67$\% of the primary beam) for each of the four pointings, this source was difficult to clean, causing artifacts across the image. We therefore peeled out the AGN from each observation by first self-calibrating on the AGN after shifting it to the phase centre to create an accurate model of the source. This model was then subtracted from the measurement set. The amplitude self-calibration was then inverted in the subtracted measurement set to reverse the effects of this off-axis amplitude self-calibration, which is likely dominated by pointing errors \citep[for an in-depth description of this technique see][]{deller15}. No further self-calibration was applied to the resulting measurement set due to the lack of suitably bright sources. The measurement sets were then imaged out to the 10\% primary beam response point using a Briggs weighting scheme of Robust=0.5, with each set of quadrant pointings then being mosaicked in the image plane, taking into account the primary beam response. The position and source flux densities of the transient described in Section~\ref{results_radio} were measured using the {\sc CASA} tool {\sc imfit} and assuming Gaussian errors based on \citet{condon98}. 

\begin{table}
\begin{center}
\caption{VLA 6\,GHz flux densities and upper-limits of the M81 transient \vtc{}}
\label{tab:vla}
\begin{tabular}{ccccc}
\\
\hline
Date & Date & Config & Quadrant & Flux density \\
(yyyy-mm-dd) & (MJD) & & & ($\mu$Jy) \\
\hline
2014-01-05 & 56662.11 & B & 1 & <23.4 \\
2014-01-06 & 56663.15 & B & 2 & <22.0 \\
2014-04-05 & 56752.03 & A & 1 & <18.3 \\ 
2014-09-12 & 56912.57 & D & 1 & <60.0 \\
2015-01-02 & 57024.28 & C & 1 & $81.0 \pm 7.7$ \\
2015-01-03 & 57025.20 & C & 2 & $112  \pm 23$  \\   
2015-03-03 & 57084.35 & B & 2 & $75   \pm 11$  \\    
2015-03-07 & 57088.01 & B & 1 & $93.9 \pm 7.8$ \\
\hline
\end{tabular}
\end{center}
Notes: The Modified Julian Date (MJD) corresponds to the midpoint of the observation. All flux density errors are $1\sigma$ and upper limits are $3\sigma$.\\
\end{table}

\section{Results: A new radio transient} \label{results}

\subsection{Radio properties} \label{results_radio}

\subsubsection{Position and luminosity} \label{results_radio_lum}

\begin{figure}
\begin{center}
\includegraphics[width=0.49\textwidth]{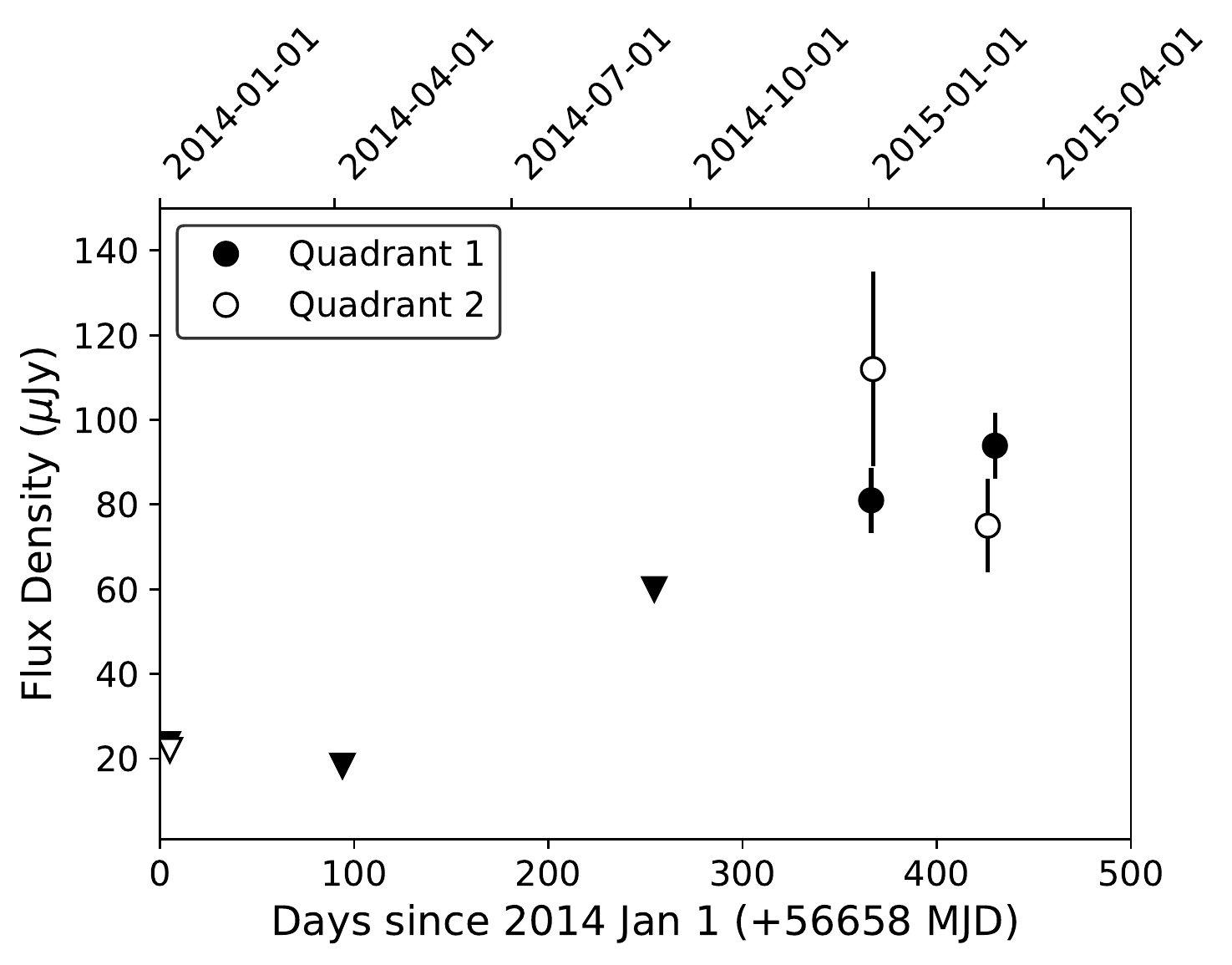}
\end{center}
\caption{The VLA 6\,GHz light curve of the M81 transient \vtc{}. The detections (circles) and $3\sigma$ upper-limits (triangles) are colour-coded to indicate which M81 quadrant was being observed at the time, with black for quadrant 1 and white for quadrant 2. Error bars are $1\sigma$.}
\label{fig:lc}
\end{figure}

\begin{figure}
\begin{center}
\includegraphics[width=0.49\textwidth]{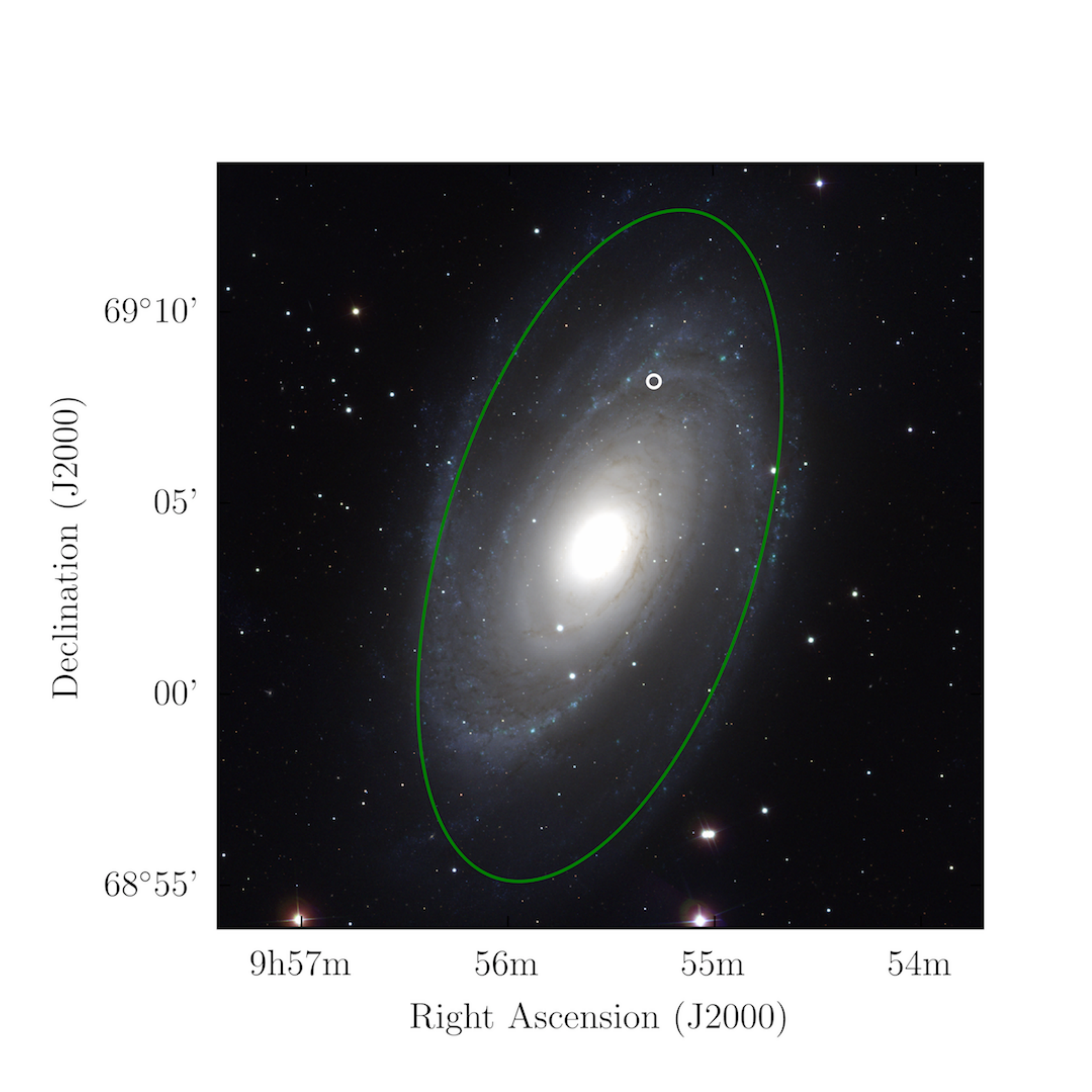}
\end{center}
\caption{An RGB image of M81 using the Sloan Digital Sky Survey filters \textit{i} (R), \textit{r} (G) and \textit{g} (B) \citep{york00,ahn12,eisenstein11}. Each band has been normalised to the 99th percentile and is displayed with an arcsinh stretch. The location of the transient \vtc{} is shown with a white circle of radius 10" ($100\times$ the $1\sigma$ error on its position). The green ellipse indicates half the major and minor diameter of the D25 (galaxy size based on the $B$ band surface brightness) for M81 \citep{devaucouleurs91}.}
\label{fig:sdss}
\end{figure}
\begin{figure*}
\begin{center}
\includegraphics[width=1.0\textwidth]{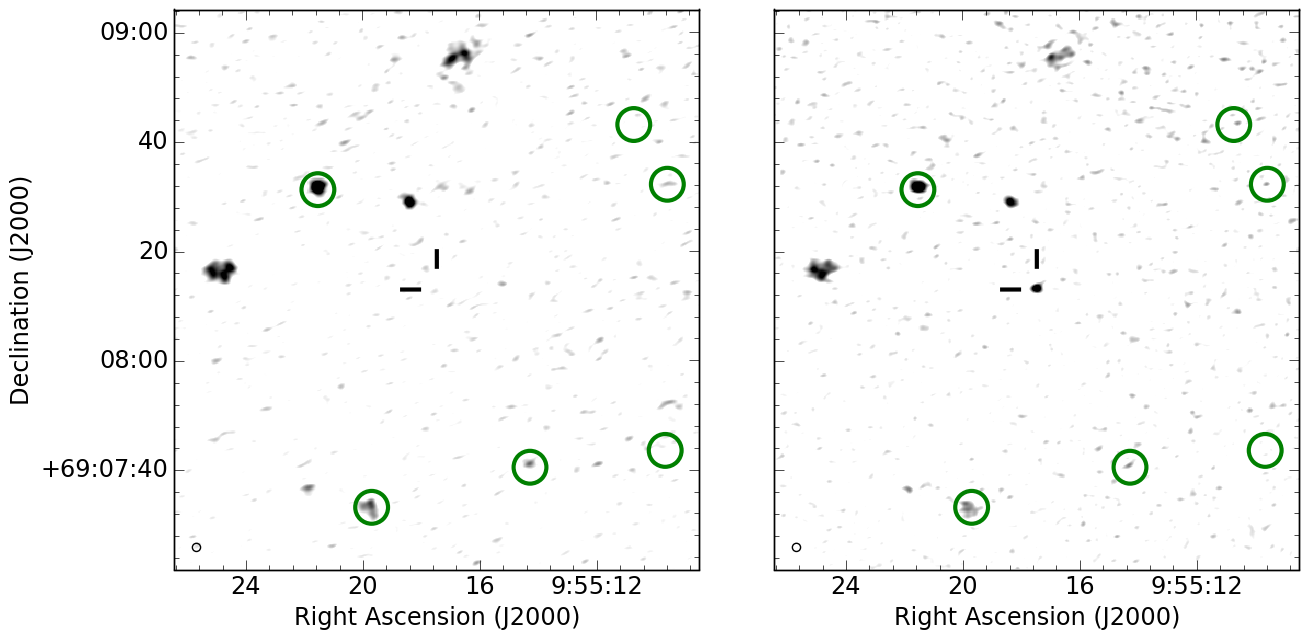}
\end{center}
\caption{B configuration mosaic observations of the region surrounding the M81 transient \vtc{}. Left: 2014 Jan mosaic when the transient was not detected. Right: 2015 March mosaic when the transient was detected. The position of the transient is indicated by the black lines. The positions of X-ray sources (signal-to-noise $>5\sigma$) detected in the stacked \textit{Chandra} data (see Section~\ref{sec:xray}) are indicated by green circles with a $3''$ radius ($\sim10$ times their position error), none of which are coincident with the radio transient.}
\label{fig:vla}
\end{figure*}

A preliminary visual inspection of several processed fields revealed a radio transient was detected in the B and C configuration observations of quadrants 1 and 2 conducted in early 2015. 
The resulting flux densities are listed in Table~\ref{tab:vla}, along with the upper limits provided by observations of quadrants 1 and 2 at earlier epochs. The VLA 6\,GHz light curve is shown in Figure~\ref{fig:lc}.
We place an upper-limit on the outburst rise-time of 112\,d, using the non-detection during the D configuration quadrant 1 observation on 2014 September 12, and the first detection of the transient in the C configuration observation of quadrant 1 on 2015 January 2. 
The transient is located within the galactic disk at the position of $\alpha \mathrm{(J2000.0)} =  09^{\mathrm{h}}55^{\mathrm{m}}17\overset{\mathrm{s}}{.}46~(\pm 0.02\mathrm{s})$ and $\delta (\mathrm{J2000.0}) = +69^{\circ}08'13\overset{''}{.}02~(\pm 0\overset{''}{.}03)$ just $4\overset{'}{.}52$ from the galactic centre of M81 (see Figure~\ref{fig:sdss}).
In quadrants 1 and 2, the transient was $\sim2\overset{'}{.}3$ and $\sim3\overset{'}{.}9$ from the pointing centres, corresponding to the 82 and 53\% response points of the primary beam, respectively. 
From hereon in we will refer to this transient as \vtc{}, 
adopting the naming convention proposed by \citet{mooley16} where ``VTC'' stands for ``VLA transient candidate". The B configuration observation both before and after the transient detection are shown in Figure~\ref{fig:vla}. 

In order to calculate the radio luminosity ($\nu L_{\nu}$) of the transient we assumed an isotropically radiating source of $\nu L_{\nu}=4 \pi d_{L}^{2} \nu S_{\nu}$, where $S_{\nu}$ is the measured flux density and $d_{L}$ is the luminosity distance equal to the Cepheid distance to M81 of $3.6 \pm 0.2$ Mpc \citep{gerke11}. As the flux densities of the transient were consistent between the observations for which it was detected, we used that of the brightest quadrant 1 detection (B configuration observed on 2015 March 7) as the transient was more centrally located within the primary beam compared to quadrant 2, resulting in a lower local root mean square (rms) noise level of $8\mu$Jy in the primary beam corrected image.
\vtc{} therefore reached a peak radio spectral luminosity of $L_{R,\nu} = 1.5 \pm 0.1 \times 10^{24}$\,erg\,s$^{-1}$\,Hz$^{-1}$ or a luminosity of $\nu L_{\nu} = 8.7 \pm 0.7 \times 10^{33}$\,erg\,s$^{-1}$ at the observing frequency of 6\,GHz. The rms at the position of the transient in this observation corresponds to a $5\sigma$ radio spectral luminosity sensitivity limit of $L_{R,\nu} \approx 6.2 \times 10^{23}$\,erg\,s$^{-1}$\,Hz$^{-1}$ (or a luminosity limit of $\nu L_{\nu} \approx 3.7 \times 10^{33}$\,erg\,s$^{-1}$ at the observing frequency of 6\,GHz) at the distance of M81. For the purpose of this paper, we will assume we are sensitive to all radio sources with an equal or higher luminosity. 

\subsubsection{Spectral index}

In order to obtain the best constraints on the spectral index of the transient over the 2\,GHz bandwidth, for quadrant 1 we stacked the visibilities from the configuration B and C detections and then imaged by cutting the observing band in half with central frequencies of 5.5 and 6.5\,GHz. This resulted in a spectral index of $\alpha = -0.5 \pm 0.8$, for $S_{\nu} \propto \nu^{\alpha}$. 
We then mosaicked this image with a stacked image of the quadrant 2, B and C configuration detections, taking into account effects from the primary beam, which resulted in a spectral index of $\alpha = -0.1 \pm 0.7$, indicating the source could be optically thick, optically thin or have components of both. All options are possible as radio transients can experience a wide variety of spectral shapes due to different contributions of optically thick and thin synchrotron emission and also bremsstrahlung \citep[e.g. core-collapse SNe and XRB radio jets;][]{weiler02,vanderlaan66,hjellming88,fender00}. 

\subsubsection{Variability}\label{radio:var}

\vtc{} was investigated for evidence of short-term variability by splitting the one hour observations with the most significant detections 
(the 2015 B and C configuration observations of quadrant 1) 
into 7 and 10 minute time bins. The flux density of the transient was measured in each of the resulting images to create a light curve and we tested for variability by calculating the $\chi^{2}$ probability that the flux density was consistent with a steady source \citep[probability $P>0.01$;][]{gaensler00}. The above analysis resulted in a minimum probability of $P=0.06$ so no significant short-term variability was detected.\footnote{Note that other point sources in the field were shown to have consistent flux densities (within $1\sigma$) between the two quadrant 1 epochs, and also showed no evidence for short-term variability.} 

This lack of short-term variability implies that the source size of \vtc{} is greater than the angular scale of interstellar scintillation, which allows us to place a lower limit on its physical source size. If we assume a scattering screen distance of 1.5\,kpc, equivalent to half the length of the Galaxy that we view through in the direction of M81, then using a scattering measure software package \citep{hancock19pp}, \footnote{https://github.com/PaulHancock/RISS19} which utilises H$\alpha$ as a proxy for the electron column \citep{haffner98} and scintillation theory by \citet{narayan92}, then our observations of M81 are in the weak scattering regime. Using formalism summarised in \citet{granot14}, we calculate a minimum source size of about $3\mu$as for \vtc{}, corresponding to a modulation index of $\lesssim0.42$ on a 1\,hr timescale. As the debiased modulation indices \citep{barvainis05} of the quadrant 1 (B and C) observations of \vtc{} are between $\sim0.1-0.2$, the source size is likely much bigger than the scintillation angular scale, corresponding to a physical source size of $\gg10$\,AU at the beginning of 2015. 
This gives a lower limit on the outburst expansion velocity of $\gtrsim150$\,km\,s$^{-1}$ assuming a maximum rise-time of 112\,days (see Section~\ref{results_radio_lum}).

The high radio luminosity observed from \vtc{} could be the result of Doppler boosting if the transient emission is from jets pointed along our line-of-sight \citep[as was the case for the M31 microquasar;][]{middleton13}. 
Minimum source size and variability arguments are often used to probe for Doppler boosting in radio transients \citep[e.g.][]{middleton13}. As no significant short-term variability was detected during the 1 hour snap-shots or between the VLA observations taken one day apart (i.e. 2015-01-02 and 2015-01-03), the most constraining lower limit we can place on the brightness temperature is using the shortest confirmed variability timescale, 
which is between the upper limit obtained on 2014 September 12, and the first detection of the transient on 2015 January 2. 
This corresponds to a change in flux of $\geq21 \mu$\,Jy in 112\,d (see Table~\ref{tab:vla}). 
As we did not resolve the rise in radio luminosity, and we know the source was not variable on timescales less than 1 hour, we can place conservative limits on the brightness temperature ($T_{\mathrm{b}}$) of $8000 \lesssim T_{\mathrm{b}} \lesssim 10^{11}$\,K, assuming our VLA observations could resolve a change in flux between 50 and 100\,$\mu$\,Jy on $\geq1$\,hr timescales \citep{readhead94}. If we assume that this is a jetted radio transient, then a comparison to the typical brightness temperature of Galactic BH XRBs \citep[$T_\mathrm{B} \lesssim 1 \times 10^{9}$\,K; see arguments in][]{middleton13}, implies a range in Lorentz factors of $0.004 \lesssim \Gamma \lesssim 5$, with the upper-limit being typical of radio transients \citep[e.g.][]{pietka15}. 

\subsubsection{Preliminary transient surface density for M81} \label{results_density}

In order to calculate the transient surface density in M81, we first calculated the total field of view surveyed. The transient search was conducted out to 50\% of the primary beam for each pointing, resulting in an area of 0.0142\,deg$^2$. By summing the areas of independent consecutive images of the same field of view, taking into account the overlap between pointings, the total area surveyed was 0.076\,deg$^2$. The detection of one transient implies a transient surface density of $13.2^{+30.2}_{-10.9}$\,deg$^{-2}$ \citep[$1\sigma$ Poisson errors;][]{gehrels86} to a $3\sigma$ flux density limit of 75\,$\mu$Jy, based on the RMS value at $50\%$ of the primary beam in the C configuration images, which were noisier than the A and B configuration images \citep[the transient surface density calculation follows the procedure described in][]{carbone16}. 
While we quote a D configuration flux density upper limit for the transient in Table~\ref{tab:vla}, this observation was confusion limited and therefore not sensitive enough to be used in this transient surface density calculation. 
If we consider the 14 months worth of VLA observations studied in this paper (see Table~\ref{tab:vla}), then the radio transient rate is $\sim0.9^{+2.0}_{-0.7}$\,yr$^{-1}$ ($1\sigma$ Poisson errors) per standard L$^{*}$ galaxy \citep[which is the case for M81; see definition described by][]{cooray05}.

\subsection{Multi-wavelength and archival investigations}

Archival images and catalogues were searched in order to identify any potential multi-wavelength transient counterparts, as well as any potential stellar, nebular or background galaxy associations at the position of \vtc{}. Extensive literature and transient database searches (including the Astronomer's Telegram (ATel)\footnote{http://www.astronomerstelegram.org/} and the Gamma-ray Coordinates Network\footnote{https://gcn.gsfc.nasa.gov/}) have not reported a transient at the position of this source in the radio, optical or X-ray bands. 

\subsubsection{X-ray results} \label{sec:xray}

Many jetted/outbursting radio transients have an X-ray counterpart so we searched the \swift\footnote{http://www.swift.ac.uk/1SXPS/}, \textit{The X-ray Multi-Mirror Mission (XMM-Newton)}\footnote{http://xcatdb.unistra.fr/3xmmdr8/} and \cxo\footnote{http://cda.harvard.edu/cscview/} point source archives at the position of \vtc{} but found no corresponding X-ray source out to 10s of arcseconds. If this source is a recurring transient, then the lack of an archival X-ray counterpart may be due to the central source being ``off" for the majority of its cycle so that current missions with sporadic coverage have not observed it during periods in outburst.

We searched for X-ray associations in all \cxo\ data taken with the Advanced CCD Imaging Spectrometer \citep{garmire03} available from the public archive, including the unpublished observations from 2016 and 2017 (PI: Swartz, Proposal Number 17620477). 
A total of 23 \cxo-ACIS observations covering the location of the transient were individually inspected. 
The \textit{Chandra} Interactive Analysis of Observations ({\sc CIAO}) software v4.10 was used for the \cxo\ data analysis, where the observations were reprocessed using {\sc CIAO} task {\sc chandra\_repro} \citep{fruscione06}. 
Table~\ref{tab:cxo} lists the measured fluxes, luminosities and 90\% upper-limits at the position of \vtc{}, which were calculated using \cxo-PIMMS\footnote{http://cxc.harvard.edu/toolkit/pimms.jsp} for count-rate conversion, assuming a column density of $n_{H} = 8 \times 10^{20}$\,cm$^{-2}$ (two times the Galactic value to account for the M81 environment) and a power law index of $\Gamma=1.8$. 
These include three possible detections with a 90\% significance but are likely to be due to fluctuations arising from small-number statistics.
However, Table~\ref{tab:cxo} also includes the fluxes and luminosities measured by stacking all \cxo\ observations, and also those from both before and after the transient ``outburst" in 2015, and represent a consistent $2\sigma$ excess.
For the purposes of this paper we will assume an upper-limit on the unabsorbed X-ray luminosity of $L_{x} \lesssim 10^{36}$\,erg\,s$^{-1}$ at the position of \vtc{}. The nearby significant ($\geq5\sigma$) X-ray point sources surrounding \vtc{} are indicated in Figure~\ref{fig:vla}.

\begin{table}
\begin{center}
\caption{\cxo\ absorbed flux and unabsorbed luminosity detections or 90\% upper-limits of the M81 radio transient \vtc{}.}
\label{tab:cxo}
\begin{tabular}{cccc}
\\
\hline
Date & ObsID$^{\mathrm{~a}}$ & Flux$^{\mathrm{~b}}$ & Luminosity$^{\mathrm{~c}}$  \\
(yyyy-mm-dd) & & (erg\,s$^{-1}$\,cm$^{-2}$) & (erg\,s$^{-1}$) \\
\hline
2000-05-07 & 735    &   $<6\times 10^{-15}$   				    &	   $< 1\times 10^{37}$        \\                                     
2005-05-26 & 5935   &   $3.8_{-3.4}^{+4.7} \times 10^{-15}$  	&	   $7.0_{-6.3}^{+9.0}\times 10^{36}$ \\
2005-05-28 & 5936   &   $<6\times 10^{-15}$     				    &	   $< 1\times 10^{37}$  \\                   
2005-06-01 & 5937   &   $<3\times 10^{-15}$      				&	   $< 6\times 10^{36}$ \\ 
2005-06-03 & 5938   &   $<4\times 10^{-15}$      				&	   $< 8\times 10^{36}$     \\         
2005-06-06 & 5939   &   $<4\times 10^{-15}$      				&	   $< 8\times 10^{36}$    \\                                
2005-06-09 & 5940   &   $<4\times 10^{-15}$       				&	   $< 7\times 10^{36}$     \\                    
2005-06-11 & 5941   &   $<6\times 10^{-15}$       				&	   $< 1\times 10^{37}$ \\                                         
2005-06-15 & 5942   &   $<2.7\times 10^{-15}$       				&      $< 5\times 10^{36}$    \\             
2005-06-18 & 5943   &   $<3.3\times 10^{-15}$        			&	   $< 6\times 10^{36}$     \\      
2005-06-21 & 5944   &   $<5\times 10^{-15}$        				&      $< 8\times 10^{36}$      \\    
2005-06-24 & 5945   &   $5.5_{-5.0}^{+5.5}\times 10^{-15}$       &	   $1.0_{-0.9}^{+1.0}\times 10^{37}$ \\
2005-06-26 & 5946   &   $<4\times 10^{-15}$       				&	   $< 7\times 10^{36}$   \\                               
2005-06-29 & 5947   &   $<5\times 10^{-15}$       				&	   $< 9\times 10^{36}$    \\     
2005-07-03 & 5948   &   $<6\times 10^{-15}$       				&	   $< 1\times 10^{37}$   \\
2005-07-06 & 5949   &   $<5\times 10^{-15}$       				&	   $< 8\times 10^{36}$   \\                                         
2016-07-04 & 18047  &   $<8\times 10^{-15}$       				&	   $< 1\times 10^{37}$  \\                         
2017-01-08 & 18048  &   $<6\times 10^{-15}$       				&	   $< 1\times 10^{37}$ \\                 
2017-01-15 & 18053  &   $<8\times 10^{-15}$       				&	   $< 2\times 10^{37}$  \\                 
2016-06-21 & 18054  &   $8.8_{-8.3}^{+11.2}\times 10^{-15}$    	&	   $1.6_{-1.5}^{+2.1}\times 10^{37}$ \\
2016-06-24 & 18875  &   $<8\times 10^{-15}$       				&	   $< 2\times 10^{37}$ \\
2017-01-11 & 19981  &   $<7\times 10^{-15}$       				&	   $< 1\times 10^{37}$  \\                             
2017-01-12 & 19982  &   $<8\times 10^{-15}$       				&	   $< 1\times 10^{37}$  \\
\hline
Before 2015 & $735-5949$ & $2.9 \pm 1.7 \times 10^{-16}$ & $5.3 \pm 3.1 \times 10^{35}$ \\
After 2015 & $18047-19982$ & $2.8 \pm 1.5 \times 10^{-16}$ & $5.0 \pm 2.8 \times 10^{35}$ \\
2000 - 2017 & All & $3.4 \pm 1.7 \times 10^{-16}$ & $6 \pm 3 \times 10^{35}$ \\
\hline
\end{tabular}
\begin{tablenotes}
\item \textit{Notes.} The first section of the table lists the detections and 90\% upper limits for individual epochs for which the position of \vtc{} falls within the \cxo\ field-of-view. All individual epoch detections quote a $1\sigma$ error. %\newline
\item The second section of the table gives the detections if stacking the \cxo\ observations that occurred both before and after the VLA detections of \vtc{} in 2015. These detections quote a $1\sigma$ error. The flux and luminosity value derived from combining all \cxo\ observations of the source are also included and are consistent with those values derived from stacking data before and after 2015. %\newline
\item $^{\mathrm{~a}}$The \cxo\ observation identification number. %\newline
\item $^{\mathrm{~b}}$Absorbed X-ray flux or 90\% upper limit at the position of the M81 transient in the $0.3-10.0$\,keV energy band. %\newline
\item $^{\mathrm{~c}}$Unabsorbed X-ray luminosity or 90\% upper limit at the position of the M81 transient in the $0.3-10.0$\,keV energy band.
\end{tablenotes}
\end{center}
\end{table}

\subsubsection{Optical and infrared results}\label{sec:oir}

We performed a catalogue search in VizieR\footnote{http://vizier.u-strasbg.fr/} for coincident optical and infrared sources. The closest reported source was detected in the infrared with the \textit{Spitzer Space Telescope}, lying $0\overset{''}{.}8$ from \vtc{} at position of $\alpha \mathrm{(J2000.0)} =  09^{\mathrm{h}}55^{\mathrm{m}}17\overset{\mathrm{s}}{.}6040$ and $\delta (\mathrm{J2000.0}) = +69^{\circ}08'12\overset{''}{.}804$ \citep{khan15}. \vtc{} is also located within (or projected on) a star-formation region with a radial extent of $\sim4{''}$, lying $1\overset{''}{.}6$ ($\sim30$\,pc at the distance of M81) from its centre at a position of $\alpha \mathrm{(J2000.0)} =  09^{\mathrm{h}}55^{\mathrm{m}}17\overset{\mathrm{s}}{.}640$ and $\delta (\mathrm{J2000.0}) = +69^{\circ}08'14\overset{''}{.}28$ \citep{hoversten11}.

\begin{figure}
\begin{center}
\includegraphics[width=0.46\textwidth]{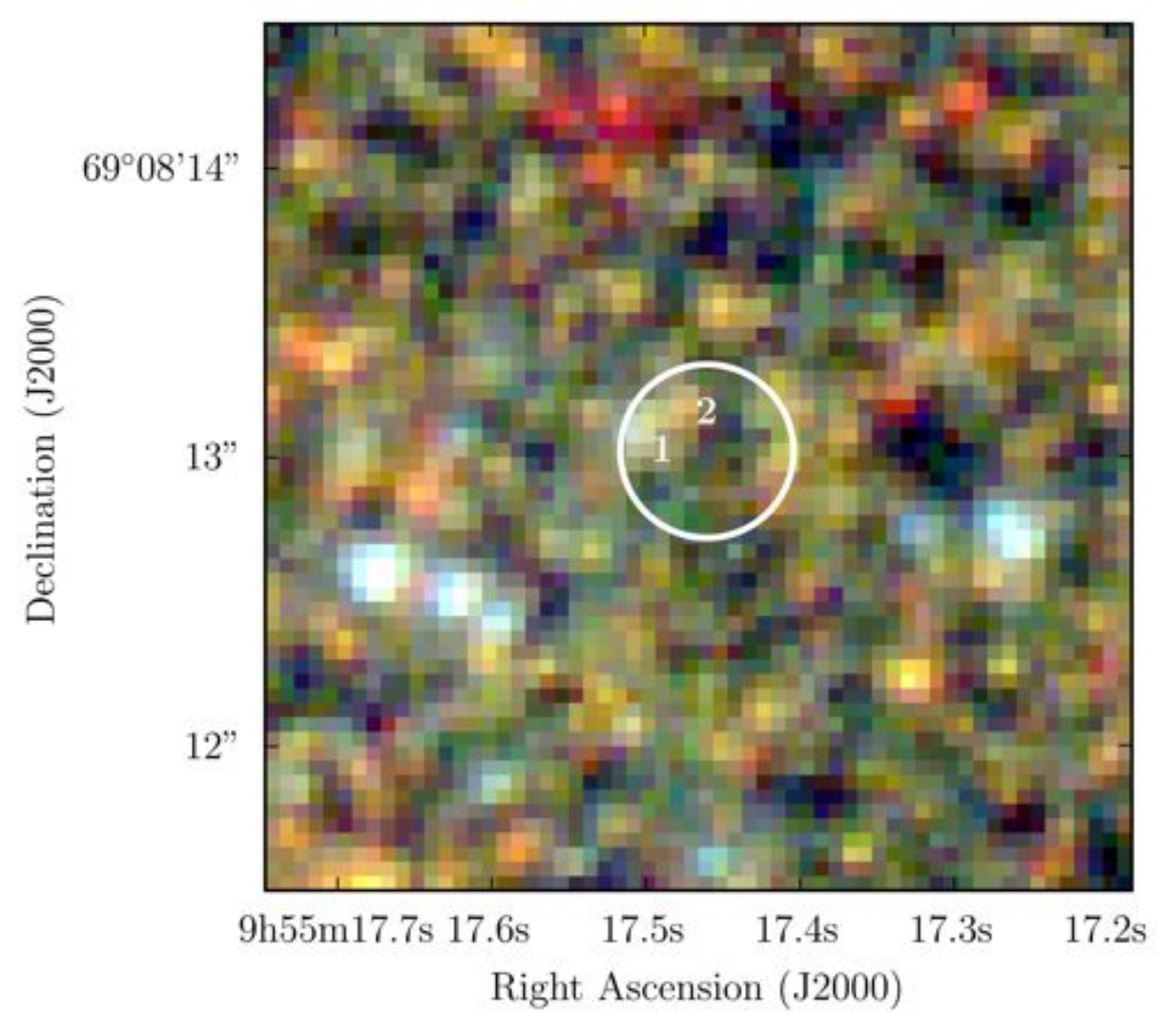}
\end{center}
\caption{Archival \hst{} image of the region surrounding the M81 transient \vtc{} (red: F814W; green: F606W; blue; F435W). Each band has been normalised to the 99th percentile and is displayed with an arcsinh stretch. The position of the transient is indicated by the white circle with a radius of 0.3'' (three times the positional error in RA, which dominates the error in Dec). The two optical sources (indicated by numbers) are investigated as possible counterparts in Section~\ref{sec:oir}.}
\label{fig:hst}
\end{figure}

In order to perform a more thorough search for an optical counterpart, we downloaded archival \hst{} data from the Hubble Legacy Archive\footnote{https://hla.stsci.edu/hlaview.html} that covered the position of \vtc{}. Note that all archival \textit{HST} observations of M81 predate the radio transient. We obtained images in each of the F435W, F606W, and F814W filters, all of which were taken with the Advanced Camera for Surveys (ACS) Wide Field Camera.\footnote{http://www.stsci.edu/hst/acs} The F435W and F606W band observations were taken on 2006-03-20 under proposal ID 10584, while the F814W observation was obtained on 2004-09-13 under proposal ID 10250. The astrometry of all three images was corrected to the Two Micron All Sky Survey (2MASS) reference frame, which has an accuracy of $0\overset{''}{.}1$ \citep{skrutskie06}. The resulting three-colour \hst\ image of the region surrounding \vtc{} is shown in Figure~\ref{fig:hst} and clearly demonstrates it resides in a crowded field. The radio position of \vtc{} is indicated by the white circle with a radius of $0\overset{''}{.}3$ ($3\sigma$ error circle; the positional error on its radio position is dominated by the error in RA of $0\overset{''}{.}1$).

Two point-like sources lie just within the positional error radius of \vtc{} as labelled in Figure~\ref{fig:hst}. We carried out aperture photometry of both sources with the Statistical Analysis package within SAOimage DS9\footnote{http://ds9.si.edu/site/Home.html}.
Their count-rates were first derived using a source radius of $0\overset{''}{.}15$ and a background annulus between $0\overset{''}{.}5$ and $1\overset{''}{.}0$ (excluding other stars), and then the revised aperture corrections{\footnote {http://www.stsci.edu/hst/acs/analysis/apcorr}} from $0\overset{''}{.}15$ to an infinite aperture were applied. 
We converted the observed net count rates to Vegamag and AB mag units using the ACS WFC zeropoints{\footnote{https://acszeropoints.stsci.edu/}} suitable for the observation dates. 
To estimate the absolute brightness of the two optical sources, first we de-reddened them, assuming Galactic line-of-sight extinction \citep{schlafly11}; the extinction coefficients for the ACS filters were computed with the York Extinction Calculator \citep{mccall04}. 
Finally, we applied a distance modulus of 27.8 mag, corresponding to our assumed distance of 3.6 Mpc, for sources located inside M81. The observed apparent brightness and de-reddened absolute brightness (both in the Vegamag and AB systems) for the two sources are listed in Table~\ref{tab:hst}.

By comparing the brightness and colors with a set of theoretical isochrones from the PARSEC v1.2 code \citep{bressan12,marigo17}, we note that if the two sources are stars inside M81, then they are consistent with red giants; they are too red to be main-sequence OB stars or blue supergiants, and too faint to be red supergiants. Thus, they are consistent with an intermediate-age population. 

Given the crowdedness of the field, we also calculated the number of optical sources likely to be randomly aligned with a $0''.3$ radius circle within this region of the M81 galaxy. By examining a $5'' \times 5''$ region centred at the position of \vtc{}, we determined there are likely to be $\sim0.6$ and $\sim2.8$ point sources that are equal to or brighter than sources 1 and 2, respectively, in the F606W and F814W bands, within the $0''.3$ radius positional error circle, with $\sim0.3$ objects equal or brighter than source 1 in the F435W band. Therefore, we conclude that sources 1 and 2 do not have any special property that distinguishes them from any other source in the field, and there is not evidence to suggest that either of them are associated with the radio transient. 

\begin{table*}
\begin{center}
\caption{\hst{} apparent and extinction corrected absolute Vega and AB magnitudes of two nearby optical point-sources within $0\overset{''}{.}3$ of the M81 radio transient \vtc{}.}
\label{tab:hst}
\begin{tabular}{ccccccc}
\\
\hline
Source & Filter & $m_{\mathrm{Vega}}$ & $m_{\mathrm{AB}}$ & $A_{\lambda}$ &$M_{\mathrm{Vega}}$ & $M_{\mathrm{AB}}$ \\
& & (mag) & (mag) & (mag) & (mag) & (mag) \\
\hline
1 & F435W & $25.8 \pm 0.1$ & $25.7 \pm 0.1$ & 0.29 & -2.3 & -2.4 \\
1 & F606W & $25.1 \pm 0.1$ & $25.2 \pm 0.1$ & 0.20 & -2.9 & -2.8 \\
1 & F814W & $24.3 \pm 0.1$ & $24.7 \pm 0.1$ & 0.12 & -3.6 & -3.2 \\
2 & F435W & $>27.0$ & $>27.0$ & 0.29 & $>-1.1$ & $>-1.1$ \\
2 & F606W & $26.2 \pm 0.2$ & $26.3 \pm 0.2$ & 0.20 & -1.8 & -1.7 \\
2 & F814W & $24.5 \pm 0.1$ & $24.9 \pm 0.1$ & 0.12 & -3.4 & -3.0 \\
\hline
\end{tabular}
\end{center}
\end{table*}

\section{Discussion: The nature of the radio transient}\label{lab:dis}

In order to determine the nature of \vtc{}, we first look at the physical properties of the radio emission. The radio flaring activity observed from slowly evolving transient sources (timescales > 1s) is thought to arise from powerful kinetic outflows interacting with the surrounding medium. This causes the compression of magnetic fields and particle acceleration, usually giving rise to synchrotron radiation.  Unfortunately, the derived spectral index of $\alpha = -0.1 +/- 0.7$ is not constraining enough to provide insight into the nature of the radio emission.

The VLA sampling of M81 did not resolve the initial rise in emission so we cannot adequately constrain its rise-time. However, recent work by \citet{pietka15,pietka17} has demonstrated that a broad correlation exists between the peak radio luminosity and the variability timescales (rise/decline rates) of a transient. Using this relation, the measured peak radio luminosity of \vtc{} ($\nu L_{\nu} = 8.7 \times 10^{33}$ erg\,s$^{-1}$) is brighter than most XRB flares, with the correlation suggesting the transient should have had an exponential rise timescale (rise-time or e-folding time) of $\tau \sim 2.6$\,d \citep[see figure 7 of][but note the scatter on the rise-time for XRBs, which is $10^{-2} \lesssim \tau \lesssim 10^{3}$ days]{pietka17}. If we assume a non-relativisitic expansion velocity, then for a maximum possible rise-time of 112\,d, the source size ($ct$) could be as large as $1.9 \times 10^{4}$\,AU by the first detection on 2015 Jan 2. 
We therefore limit the size of \vtc{} to be between 10\,AU (from scintillation; Section~\ref{radio:var}) and $1.9 \times 10^{4}$\,AU.

Further physical properties of the emitting plasma can be derived from the luminosity and rise-time of the radio ejection if we assume it is synchrotron in nature. Using the potential rise-time range from 2.6 to 112 days for \vtc{} and a peak flux of 100\,$\mu$Jy, we calculate the minimum energy of the plasma (assuming the source expands at $c$) to be $10^{44} \lesssim E_{\mathrm{min}} \lesssim 10^{46}$\,erg, the minimum (jet) power associated with the ejection to be $P_{\mathrm{min}} \sim10^{39}$\,erg\,s$^{-1}$, and a corresponding magnetic field at equipartition of $2-50$\,mG \citep{fender06,longair11}. 
Given the ambiguity in the rise time, and the diverse nature of radio transient timescales and brightnesses, in the following we discuss possible classifications for \vtc{}. 

\subsection{Foreground or background event}

We first explore the possibility that \vtc{} is either a foreground event within the Milky Way or a  background event observed through the M81 galactic disk. 
Table~\ref{tab:bkgfor_rates} summarises the surface density rates, timescales and luminosities of potential contaminating foreground and background radio transient populations in the area of M81 surveyed by the VLA. The surface density rates are scaled to the survey $3\sigma$ flux density threshold of $S_{\nu} \sim 75 \mu$Jy (Section~\ref{results_density}) assuming that it is a Euclidean source population that scales according to $\rho \propto S_{\nu}^{-1.5}$.
The Galactic rate combines foreground radio transients with similar timescales to \vtc{} (weeks to years), which include  XRBs, young stellar objects, pulsar scattering, magnetars and novae \citep{mooley16}. Even taking into account upper limits, the low rates of Galactic transient rates, combined with the lack of an optical association in the \hst\ data, makes it unlikely that \vtc{} is a foreground event located within the Milky Way.

Individual extragalactic radio transient populations that evolve on similar timescales to \vtc{} (weekly to 1 yr timescales) are also considered in Table~\ref{tab:bkgfor_rates}, including AGN, jetted and off-axis tidal disruption events (TDEs) \citep{mooley16}, gamma-ray burst (GRB) afterglows \citep[for typical radio timescales and luminosity ranges see][]{chandra12,anderson18}, and off-axis (orphan) GRBs \citep{law18}.
A comparison of the extragalactic rates indicate that AGN are the most likely contaminants, with a potential contribution of up to one within the total area surveyed with the VLA.
This is reasonable consistent with the surface density calculated by \citet{bell15} from a 5.5\,GHz survey of the \textit{Chandra} Deep Field South ($\rho <7.5$\,deg$^{-2}$ above $70 \mu$\,Jy) on 2.5 month and 2.5\,y timescales, with the 2.5 month cadence being roughly similar to a subset of timescales probed by our VLA epochs. This rate predicts $<0.5$ background transient events within the 0.076\,deg$^{2}$ area of M81 surveyed by the VLA (again making the same flux density scaling assumption).

\begin{center}
\begin{table*}
\caption{Surface densities, timescales and luminosities for foreground and background radio transients}
\label{tab:bkgfor_rates}
\begin{tabular}{lccccc}
\\
\hline
Type & Rate$^{\mathrm{~a}}$ & Rate$^{\mathrm{~b}}$ & Timescales$^{\mathrm{~c}}$ & Peak Luminosity & Ref$^{\mathrm{~d}}$\\
& (deg$^{-2}$) & (M81) & & (erg\,s$^{-1}$\,Hz$^{-1}$) \\
\hline
\hline
Galactic & 1.3 & $(5-100)\times10^{-3}$ & weeks - year & $10^{15}-10^{21}$ & 1 \\
\hline
AGN & 10.0 & 0.8 & days - years & $10^{27}-10^{34}$ & 1 \\
TDE & 0.12 & 0.01 & years & $10^{30}-10^{31}$ & 1 \\
GRB & $9\times10^{-3}$ & 0.001 & days - months & $10^{27}-10^{32}$ & 2,3 \\
Orphan GRB & $(5-100)\times10^{-2}$ & $(5-100)\times10^{-3}$ & years & $\sim10^{29}$ & 4,5,6 \\
\hline
\end{tabular}
\begin{tablenotes}
\item \textit{Notes.} The Galactic transients include contributions from XRBs, young stellar objects, pulsar scattering, magnetars and novae. The AGN included in this rate are those expected to have variability processes that evolve on similar timescales to \vtc{}, including `shock-in-jet' and extreme scattering events. The TDE rates include jetted and off-axis TDEs. The GRB and orphan GRB rates are in deg$^{-2}$\,yr$^{-1}$. The orphan GRB afterglow luminosity is based on that of the candidate FIRST J141918.9+394036, which was identified by \citet{law18} \citep[see also][]{marcote19pp}. %\newline
\item $^{\mathrm{~a}}$The surface density rate for the corresponding transient type.%\newline
$^{\mathrm{~b}}$The number of transients of this type expected in the 0.076\,deg$^{2}$ area of M81 surveyed in this paper.\newline
$^{\mathrm{~c}}$The full evolutionary timescale of the corresponding transient type.\newline
$^{\mathrm{~d}}$References - 1: \citet{mooley16}, 2: \citet{chandra12}, 3: \citet{ghirlanda13}, 4: \citep{gal-yam06}, 5: \citet{ghirlanda14}, 6: \citet{law18}
\end{tablenotes}
\end{table*}
\end{center}

While it is not possible to rule out a background variable AGN for \vtc{}, its position within the northern spiral arm of M81 (see Figure~\ref{fig:sdss}) strongly hints at an association local to M81. Additionally, given that there is no obvious optical association at the position of \vtc{} (see Figure~\ref{fig:hst}), it is unlikely that we are observing a background active galaxy through the disk of M81.
For the rest of this paper we will focus on non-nuclear radio transients that are local to M81 as possible explanations for \vtc{}. 

\pagestyle{empty}
\begin{landscape}
\begin{center}
\begin{table}
\caption{Properties of \vtc{} compared to other transient classes and sources assuming they are located in M81. The tabulated values are based on best-studied sources.}
\label{tab:class}
\begin{tabular}{lccccccccll}
\\
\hline
Source/Type & \multicolumn{2}{c}{Radio flare timescales$^{\mathrm{~a}}$} & Peak Lum & Radio short-term & X-ray & Optically & Ref$^{\mathrm{~e}}$ & Candidate$^{\mathrm{~f}}$ & Conflicting & Compatible \\
\cline{4-4}
& (Rise) & (Decay) & (erg\,s$^{-1}$\,Hz$^{-1}$) & variability$^{\mathrm{~b}}$ & detected$^{\mathrm{~c}}$ & detected$^{\mathrm{~d}}$ & & & properties$^{\mathrm{~g}}$ & properties$^{\mathrm{~h}}$ \\
 & (1) & (2) & (3) & (4) & (5) & (6) & & & & \\
\hline
\hline
\vtc{} & $2.6-112$\,d & $>2$\,months & $1.5\times10^{24}$ & N & N & N & 1 & & \\
\hline
CCSN & days - months & days - years$\dagger$ & $10^{26}-10^{29}$ & S & S & Y & 2 & N & 3,4,5,6 & 1,2 \\
Peculiar SN \textit{(Type IIP)} & days & days - years$\dagger$ & $4 \times 10^{23} - 10^{26}$ & S & S & Y & 3-5 & ? & 4,5,6 & 1,2,3 \\
Magnetar giant flare & days & days - months & $\lesssim10^{22}$ & N & Y & N & 6-8 & N & 3,5 & 1,2,4,6 \\
BNS merger \textit{(GW170817)} & months & months - years & $2\times10^{26}$ & N & Y & Y & 9-12 & N & 3,5,6 & 1,2,4 \\
LMXB & hours - days & days - months & $\lesssim10^{23}$ & Y & Y & S & 13-25 & N & 3,4,5,6 & 1,2 \\
HMXB \textit{(Cyg X-3)} & hours - days & days - months & $\lesssim 10^{24}$ & Y & Y & Y & 26-29 & N & 4,5,6 & 1,2,3 \\
Absorbed HMXB \textit{(SS 433)} & hours - days & days & $\lesssim 10^{23}$ & Y & N & Y & 30-34 & N & 2,3,4,6 & 1,5 \\
IMBH \textit{(HLX-1)} & days & days & $6.5 \times 10^{26}$ & Y & Y & Y & 35-36 & N & 2,3,4,5,6 & 1 \\
\hline
\end{tabular}
\begin{tablenotes}
\item \textit{Notes.} Some SNe go undetected in the optical band due to line-of-sight absorption.% \newline
\item $\dagger$ - Radio flare decay timescales of years mainly pertain to interacting SNe (see Section~\ref{sec:sn}). 
%\newline
\item The LMXB characteristics are based on those sources that have the brightest radio flares, including GRS 1915+105, GRO J1655-40, XTE J1748-288, H1743-322, and V404 Cyg.
%\newline
\item The HMXB characteristics are based on Cyg X-3, which has the brightest radio flares of any accreting system in our Galaxy, and one of the few such systems that would be detectable at the distance of M81 in our VLA data.
%\newline
\item The absorbed HMXB characteristics are based on SS 433, which is the only such system in our Galaxy.
%\newline
\item The IMBH properties are based on HLX-1, which is the strongest IMBH candidate that also produces radio flares. 
%\newline
\item $^{\mathrm{~a}}$The typical radio flare timescales for both the rise and decay phases.
%\newline
\item $^{\mathrm{~b}}$Intrinsic short-term radio variability on timescales shorter than a day where Y = Yes, N = No and S = Sometimes. Note that radio transients may scintillate when the source size is small. 
%\newline
\item $^{\mathrm{~c}}$X-ray detection refers to an X-ray luminosity of $L_{x}\gtrsim10^{37}$\,erg\,s$^{-1}$ where Y = Yes, N = No and S = Sometimes.
%\newline
\item $^{\mathrm{~d}}$Optical detection refers to any reported optical counterpart where Y = Yes, N = No and S = Sometimes.
%\newline
\item $^{\mathrm{~e}}$References - 1: This paper, 2: \citet{perez-torres15}, 3: \citet{turtle87}, 4: \citep{pooley02}, 5: \citep{beswick05}, 6: \citet{frail99}, 7: \citet{gaensler05}, 8: \citep{gelfand05}, 9: \citet{hallinan17}, 10: \citet{mooley18a}, 11: \citet{mooley18b}, 12: \citet{dobie18}, 13: \citet{han92}, 14: \citet{mirabel94}, 15: \citet{hjellming95}, 16: \citet{fender97}, 17: \citet{pooley97}, 18: \citet{fender99}, 19: \citet{hjellming00}, 20: \citet{fender04rev}, 21: \citet{brocksopp07}, 22: \citet{mcclintock09}, 23: \citet{jonker10}, 24: \citet{tetarenko17}, 25: \citet{tetarenko19}, 26: \citet{gregory72}, 27: \citet{waltman95}, 28: \citet{millerjones04}, 29: \citet{corbel12}, 30: \citet{fiedler87}, 31: \citet{trushkin06}, 32: \citet{trushkin11}, 33: \citet{trushkin12}, 34: \citet{trushkin18}, 35: \citet{webb12}, 36: \citep{cseh15}
%\newline
\item $^{\mathrm{~f}}$Candidate refers to whether the corresponding source type is a viable classification for \vtc{} where N = No and ? = possible.
%\newline
\item $^{\mathrm{~g}}$Characteristics that are not consistent with \vtc{}. The listed numbers correspond to the number labelling of the columns.
%\newline
\item $^{\mathrm{~h}}$Characteristics that are consistent with \vtc{}. The listed numbers correspond to the number labelling of the columns.
\end{tablenotes}
\end{table}
\end{center}
\end{landscape}

\begin{landscape}
\begin{table}
\begin{center}
\caption{Comparison of \vtc{} to Galactic LMXBs and extragalactic transients.}
\label{tab:extran}
\begin{tabular}{lccccccccccc}
\\
\hline
Source name & \multicolumn{2}{c}{Radio flare timescales$^{\mathrm{~a}}$} & Peak Radio Lum & Radio short-term & T$_{\mathrm{B,min}}^{\mathrm{~c}}$ & E$_{\mathrm{min}}^{\mathrm{~d}}$ & P$_{\mathrm{min}}^{\mathrm{~e}}$ & B$^{\mathrm{~f}}$ & Peak X-ray Lum$^{\mathrm{~g}}$ & Optically$^{\mathrm{~h}}$ & Ref$^{\mathrm{~i}}$\\
& (Rise) & (Decay) & (erg\,s$^{-1}$\,Hz$^{-1}$) & variability$^{\mathrm{~b}}$ & (K) & (erg) & (erg\,s$^{-1}$) & (G) & (erg\,s$^{-1}$) & Detected & \\
\hline
\hline
\vtc{} & $2.6-112$\,d & $>2$\,months & $1.5 \pm 0.1 \times10^{24}$ & N & $8000 - 10^{11}$ & $\sim10^{44}-10^{46}$ & $\sim10^{39}$ & $2\times10^{-3}-0.05$ & $\lesssim 6 \pm 3 \times 10^{35}$ & N & 1-2 \\
\hline
M31 microquasar & hours - days & days & $2.70 \pm 0.06 \times 10^{23}$ & Y & $7 \times 10^{10}$ & $\sim10^{44}$ & $\sim10^{38}$ & $\sim7\times10^{-3}$ & $1.26 \pm 0.2 \times 10^{39}$ & N & 3 \\
43.78+59.3 (in M82) & $\lesssim4$\,d & $>19$\,months & $\sim1.4 \times 10^{25}$ &  N & $2 \times 10^{8}$ & $\sim6 \times 10^{44}$ & $\sim10^{39}$ & $\sim0.07$ & $\lesssim 2 \times 10^{37}$ & Y$^{\dagger}$ & 4-7 \\
Ho II X-1 & ? & $\gtrsim1.5$\,y & $4.1 \pm 0.2 \times10^{24}$ & N & $1.5 \times 10^{5}$ & $2.6 \times 10^{49}$ & $10^{39} - 10^{40}$ & $\sim7.7 \times 10^{-4}$ & $\sim10^{40}$ & Y$^{\ddagger}$ & 8-14\\
\hline
\end{tabular}
\begin{tablenotes}
\item \textit{Notes.} The brightness temperature, estimated minimum energy, mean power (jet power) and magnetic field were calculated for the M31 microquasar, 43.78+59.3 and Ho II X-1 based on equations from \citet{middleton13} and \citet{fender06} if unavailable in the literature. The Ho II X-1 properties relate to the core radio emission (core-lobe morphology) rather than from the large scale radio bubble. %\newline
\item $^{\dagger}$ - A very faint optical counterpart to 43.78+59.3 was detected in NIR \citep{mattila13}. 
%\newline
\item $^{\ddagger}$ - The optical detection of Ho II X-1 relates to its nebula \citep{pakull02,kaaret04}.
%\newline
\item $^{\mathrm{~a}}$The typical radio flare timescales for both the rise and decay phases.
%\newline
\item $^{\mathrm{~b}}$Intrinsic short-term radio variability on timescales shorter than a day where Y = Yes and N = No.
%\newline
\item $^{\mathrm{~c}}$The reported brightness temperature is calculated using the shortest observed radio variability.
%\newline
\item $^{\mathrm{~d}}$The estimated minimum energy associated with a finite volume of synchrotron emitting plasma based on the peak radio luminosity and assuming equipartition.
%\newline
\item $^{\mathrm{~e}}$The related mean power (jet power) of the radio flare.
%\newline
\item $^{\mathrm{~f}}$The corresponding magnetic field strength.
%\newline
\item $^{\mathrm{~g}}$The peak X-ray luminosity in the $0.3-10.0$\,keV energy range. 
%\newline
\item $^{\mathrm{~h}}$Optical detection refers to any reported optical counterpart where Y = Yes and N = No.
%\newline
\item $^{\mathrm{~i}}$References - 1: This paper, 2: \citep{pietka17}, 3: \citep{middleton13}, 4: \citet{muxlow10}, 5: \citet{joseph11}, 6: \citet{gendre13}, 7: \citet{mattila13}, 8: \citet{zezas99}, 9: \citet{pakull02}, 10: \citet{kaaret04}, 11: \citet{miller05}, 12: \citet{cseh12}, 13: \citet{cseh14}, 15: \citet{cseh15}.
\end{tablenotes}
\end{center}
\end{table}
\end{landscape}
\pagestyle{plain}

\subsection{A transient local to M81}

The types of radio transients that are commonly observed in nearby galaxies include core-collapse SNe and some extreme XRBs. Less common potential classes may include magnetar flares and binary neutron star (BNS) mergers, though few such cases have been studied. We compare the timescale, luminosity and multi-wavelength properties of \vtc{} to these source classes in Table~\ref{tab:class}. We immediately rule out several potential classifications for \vtc{} based on a comparison with their typical peak radio luminosities. These include the rare radio flares associated with giant $\gamma$-ray magnetar flares \citep[e.g. SGR1900+14 and SGRB1806-20;][]{frail99,gaensler05,gelfand05}, which are at least 200 times less luminous than \vtc{}, making them an unlikely candidate even if they do have similar flare timescales and multi-wavelength properties. In addition, the radio emission from the BNS merger GW170817 associated with GRB 170817A \citep{abbott17a,abbott17b} also has similar timescales to \vtc{}, but with a peak luminosity two orders of magnitude brighter. The radio emission from GW170817 is thought to arise from an off-axis relativistic jet \citep{mooley18b} and is also detected across the entire electromagnetic spectrum, unlike \vtc{}, making a BNS merger scenario similarly unlikely \citep[note that due to the small sample size of gravitational-wave-detected BNS mergers, we do not know the full range of their energetics, particularly for more off-axis or choked jet scenarios; see][for a summary]{mooley18a}. We discuss our comparisons to SNe and XRB accreting systems in more detail in the following.

\subsubsection{Radio Supernova}\label{sec:sn}

M81 is regularly monitored for novae, yielding several detections per year \citep{hornoch08}, which are routinely reported as ATels but no optical transient was reported at the position of \vtc{}. This does not necessarily rule out a SN identification as an archival search at 1.4\,GHz resulted in the discovery of a heavily obscured, presumably Type II (core-collapse) SN in galaxy NGC 4216 \citep{levinson02,gal-yam06}.
However, this scenario is unlikely given that low extinction levels are expected along the line-of-sight when viewing a face-on spiral. Therefore, the lack of any reported optical transient at the position of \vtc{} already discourages a SN association.

A radio luminosity and timescale comparison of \vtc{} to core collapse (CC) (including peculiar) SNe can be found in Table~\ref{tab:class}. 
While CC SNe span a wide range of peak radio spectral luminosities \citep{perez-torres15},\footnote{Note that thermonuclear Type Ia SNe remain undetected in the radio band \citep[$L_{R,\nu} < 3 \times 10^{23}$\,erg\,s$^{-1}$\,Hz$^{-1}$;][]{chomiuk16}.} the peak luminosity of \vtc{} is at least an order of magnitude fainter than most known events, with the exception of the peculiar (Type IIP) SN 1987A, which peaked at $L_{R,\nu} = 4 \times 10^{23}$~erg s$^{-1}$ Hz$^{-1}$ but faded away within days \citep{turtle87}. 
However, SN 1987A is also an interacting SN as it has shown a continuous increase in radio flux for over 25 years \citep{zanardo10,cendes18} and is now exceeding the initial reported peak luminosity. This late-time radio emission is produced by a collisionless shock that is generated by the deceleration of the SN ejecta by circumstellar material \citep[e.g.][]{katz10,murase11,murase14}. In fact, many radio SNe can be extremely long-lived \citep[$>10$\,y, e.g. SN 1988Z;][]{williams02}, show late-time light curve features \citep[e.g. SN 2014C;][]{anderson17}, and in some cases sinusoidal modulations due to companion interactions \citep[e.g. SN 1979C, 2001ig;][]{weiler92,ryder04}. 
Given the longevity of \vtc{} ($>2$\,months), we cannot rule out radio emission from late-time shockwave interactions with the circumstellar medium produced by a CC SNe at luminosities similar to SN 1987A.

\subsubsection{X-ray binary or IMBH}

Table~\ref{tab:class} also shows a comparison between the peak radio luminosities and decay timescales of \vtc{} and XRBs.
We rule out high-mass X-ray binaries (HMXBs) as they have bright optical counterparts (usually from an OB companion star), whereas \vtc{} does not (see Figure~\ref{fig:hst}). In addition, the only Galactic HMXB that even reaches similar radio luminosities to \vtc{} is Cyg X-3 but usually over much faster timescales, with the jets displaying short-term variability \citep{gregory72,waltman95,millerjones04,corbel12}. Unlike \vtc{}, Cyg X-3 is also always X-ray bright, with average X-ray luminosities ranging from $L_{x} \sim 10^{37} - 10^{39}$\,erg\,s$^{-1}$ \citep[e.g.][and references therein]{watanabe94,szostek08,hjalmarsdotter08,koljonen10}. For similar reasons we also rule out an intermediate mass black hole (IMBH; $\sim10^{2}-10^{5}$ $M_{\odot}$) accreting system analogous to HLX-1 \citep{farrell09}, which is also optically and X-ray luminous, with rapidly flaring radio jets that are $\sim400$ times more luminous than \vtc{} \citep{webb12,cseh15}.

Galactic low-mass X-ray binaries (LMXBs) are known to emit flaring radio jets as they transition between different X-ray spectral or accretion states, particularly from the hard to the soft X-ray state \citep{fender04}. Even the most powerful LMXBs in the Milky Way cannot produce radio flares that are detectable out at M81 at the sensitivity of our VLA observations, with the majority of flares only lasting a few days (see Table~\ref{tab:class}). 
Nonetheless, there are a few cases of Galactic LMXBs that have experienced radio flares with much longer decay phases \citep[$\sim100$\,days, e.g. V404 Cyg, V6461 Sgr and H1743-322;][]{han92,hjellming00,mcclintock09} or periods of stable radio flux \citep[GRS 1915+105;][]{fender04rev}, which are more consistent with the timescale of \vtc{}.
However, typical LMXB minimum energies are also at least an order of magnitude lower than our estimate for \vtc{} (see Table~\ref{tab:extran}), with few reaching a similar jet power \citep{pietka17,fender99,brocksopp07,curran14,tetarenko17,dunn10}.
Another important difference to LMBXs is the lack of short-term radio variability displayed by \vtc{} within the four 1 hour epochs at which \vtc{} was detected.
The lack of X-ray counterpart and short term radio variability, combined with the longer decay timescales and high radio luminosity argue against an LMXB scenario for \vtc{}.

The other form of accreting system worth comparing to are absorbed Galactic HMXBs such as SS 433, which has radio flares on the order of 1-2\,Jy \citep{fiedler87}, but a faint X-ray luminosity \citep[$L_{x} \sim 10^{35}-10^{36}$\,erg\,s$^{-1}$;][]{marshall79,marshall02}, when compared to other HMXBs. 
Its X-ray faintness is likely due to the central engine being viewed edge-on and thus obscured by material along the line-of-sight \citep[note that recent observations with the \textit{Nuclear Spectroscopic Telescope Array} demonstrate that SS 433 is likely accreting in the super-Eddington regime;][]{middleton18pp}. 
However, the radio flares from SS 433 are shorter lived, less luminous \citep[e.g.][]{trushkin06,trushkin11,trushkin12,trushkin18} and experience a shallower dynamic range between their peak and the quiescent radio luminosity when compared to \vtc{} \citep{fiedler87}. 
Nonetheless, SS 433 also has persistent radio emission produced by its powerful jets impacting the surrounding environment, causing them to inflate an elongated synchrotron nebula \citep[bubble; e.g.][]{dubner98,broderick18}.

Overall, the longevity of \vtc{} is still at odds with all the Galactic accreting systems we have discussed thus far and is therefore unusual when compared to what we usually observe from radio jets associated with accretion powered outbursts. We therefore turn our attention to the more elusive extragalactic radio transients, comparing \vtc{} to those that have been identified as XRBs or at least accretion candidates.

\subsection{Comparison to extragalactic accreting transients}

A small number of non-nuclear and likely accreting extragalactic radio transients have been reported in the literature. These include the transients detected in the M31 \citep{middleton13}, M82 \citep{muxlow10}, and Cygnus A \citep{perley17} galaxies. We immediately disregard the Cygnus A transient as it is five orders of magnitude more radio luminous ($L_{R,\nu} = 3 \times 10^{29}$\,erg\,s$^{-1}$\,Hz$^{-1}$) than \vtc{} and is suggested to be a secondary supermassive black hole in orbit around the Cygnus A primary. In Table~\ref{tab:extran} we compare the flare timescales, peak radio luminosity, short timescale variability, multi-wavelength characteristics and properties such as the brightness temperature, minimum energy, jet/minimum power and magnetic field strength \citep{fender06} of extragalactic radio transients to \vtc{}.

Table~\ref{tab:extran} shows that \vtc{} has quite different characteristics to the M31 microquasar, which was one of the most radio luminous extragalactic XRBs ever detected. This microquasar also showed short-term variability on minute timescales that could be attributed to either intrinsic variations or interstellar scintillation, which demonstrated its unusually high radio luminosity was due to the jet being Doppler boosted \citep{middleton13}. While \vtc{} is a factor of $\sim5$ times more radio luminous than the M31 transient and at least that many times more luminous than the brightest Galactic XRBs, it showed no evidence for short-term variability (and therefore scintillation), indicating the radio source was unlikely to be compact. However, this does not discount variability for \vtc{} if compared to a random sampling of the M31 transient radio detections. If placed at the distance of M81, the M31 transient could look stable over the $30-40$ days of radio monitoring (though barely detected at $3\sigma$ in our VLA observations). The M31 microquasar was also X-ray bright and briefly accreting in the Eddington regime.

A characteristic that both the M31 and M81 transients do share is a lack of a bright optical counterpart. \citet{middleton13} argue that this rules out a high-mass OB-type donor star and that the M31 transient is more likely to be accreting from a low-mass companion. If \vtc{} is an accreting system, it too may have a low mass donor star. A low donor mass also prevents us from drawing associations with nearby star-forming regions as LMXBs receive a natal kick from the SN of the primary, resulting in peculiar velocities of between tens and a few hundred km\,s$^{-1}$ \citep{miller-jones14}.

The M81 radio transient shares more characteristics with 43.78+59.3, the radio transient in M82 that was discovered by \citet{muxlow10}. Several classifications were suggested but the most likely is a stellar mass accreting system \citep{muxlow10,joseph11}. However, 43.78+59.3 would be among the brightest microquasar radio flares detected, being a factor of 10 times more luminous than \vtc{} \citep[as there are only approximate distance estimates for M82, e.g.][we assume the M81 Cepheid distance to M82 as these two galaxies are interacting]{dalcanton09,foley14}. 
In order to investigate the effect of Doppler boosting on the observed radio luminosity of 43.78+59.3, we used the shortest observed variability, which was between two observations separated by $\sim4$ days when the source was observed to switch-on \citep{muxlow09,brunthaler09,muxlow10}, with the flux density change corresponding to a brightness temperature of $T_\mathrm{B} \gtrsim 2 \times 10^{8}$\,K, which is not constraining enough to confirm beaming (see arguments in Section~\ref{radio:var}). 43.78+59.3 also shows a minimum energy, related mean jet power and magnetic field strength values consistent with those calculated for \vtc{} (see Table~\ref{tab:extran}).

The potential identification of 43.78+59.3 as a traditional LMXB was challenged by \citet{joseph11} as it displayed atypical characteristics, such as its longevity and lack of short-term variability, with a nearly constant optically thin radio spectral index of $\alpha \sim -0.7$ over more than 150\,d \citep{muxlow10}. Combined with no detectable X-ray counterpart, \citet{joseph11} suggested that 43.78+59.3 may be an SS 433 analogue due to similarities such as its more consistent radio flux density compared to other microquasar flares, its optically thin (steep) radio spectrum \citep[$\alpha \approx -0.5$;][]{dubner98} and its faint X-ray luminosity. An extragalactic analogue of SS 433 is NGC 7793-S26 \citep{soria10}, again showing radio emission in the form of lobes and a surrounding cocoon, rather than from the core. 
Additional evidence for a radio nebular identification for 43.78+59.3 is that it was marginally resolved by very long baseline interferometry (VLBI) observations when it was 250\,d old, which indicated an expansion velocity that would not be possible for a SN \citep{muxlow10}. 

The properties of \vtc{}, such as its luminosity, longevity and lack of short-term variability, argue against a transient jet like those seen from Galactic and extragalactic XRBs (such as the M31 microquasar) but may still indicate the interaction of XRB jets with the surrounding environment. We now investigate the possibility of \vtc{} being an unresolved radio nebula associated with an outbursting accretion event.

\subsection{The birth of a ULX radio nebula}

The M82 transient 43.78+59.3 and potentially \vtc{} show properties that may be indicative of a newly born synchrotron radio nebula and thus similar to those of ULX radio bubbles and/or lobes. Traditionally, a ULX classification is based on the X-ray luminosity threshold of $L_{x} \gtrsim10^{39}$\,erg\,s$^{-1}$. 
However, there are several sources that have a radiative luminosity below this limit but have a kinetic power inferred from their surrounding nebula (jet power) of $P_{\mathrm{jet}} > 10^{39}$erg\,s$^{-1}$ \citep[e.g. SS 433, NGC 7793-S26, M83 MQS1, and NGC 300-S10;][]{marshall02,pakull10,soria10,soria14,urquart19}. 
The low levels or lack of X-ray emission could be due to the source being in a low accretion state or our view being obscured by absorbing material along our line-of-sight (e.g. SS 433). For example, the source may be viewed along the plane of the accretion disk, which is likely puffed up in its outer regions, and additional obscuration could be caused by a circumbinary disk in the same plane \citep{kaaret17}.
Nonetheless, such a high jet power indicates periods of super-critical accretion regardless of whether the radiative power of the source is observed to be sub-Eddington. As we are interested in jet power for the purposes of this study, we will refer to those microquasars with super-critical jet powers inferred from their surrounding nebulae as ULXs (i.e. those listed above).

Both \vtc{} and 43.78+59.3 are X-ray faint but have radio luminosities consistent with those of ULX radio bubbles \citep[$\nu L_{\nu} \sim 2 \times 10^{34}$ to $2 \times 10^{35}$\,erg\,s$^{-1}$ at 5\,GHz;][]{vandyk94,miller05,urquhart18} and as previously mentioned, \citet{joseph11} have already suggested that 43.78+59.3 is an analogue of SS 433 and NGC 7793-S26. \footnote{Note that radio nebulae produced by ULXs and Galactic XRBs are uncommon as their existence is dependent on the density of the surrounding medium, timescales of the energy injection, and the jet power. A low density environment causes any potential radio lobes to quickly dissipate \citep{heinz02}.}
Methods for calculating the jet power from the radio luminosity of a microquasar nebula have proven to be very inaccurate so we instead used the peak luminosity, assuming equipartition, to calculate a mean power of $\sim10^{39}$\,erg\,s$^{-1}$ for \vtc{} and 43.78+59.3, indicating that both may be consistent with a super-Eddington accreting source (see Table~\ref{tab:extran}).

One big difference between \vtc{}, 43.78+59.3 and existing ULX radio bubbles are their lifetimes. ULX bubbles have characteristic lifetimes of $\sim10^{4}$ to $10^{5}$\,years) and such radio nebulae are usually assumed to be powered by nearly continuous energy injection for over that period \citep{soria10}. 
However, high spatial resolution radio observations of Ho II X-1 have demonstrated that there is a triple radio structure within its larger radio bubble. This core and lobe morphology is indicative of a collimated jet structure produced by discrete ejecta \citep{cseh14}. All three radio components are optically thin, arguing against the direct detection of a self-absorbed compact jet usually associated with the canonical hard state of Galactic XRBs but instead suggesting that the lobes are from a former epoch of activity while the core, which is brighter than the lobes, is associated with the most recent ejection event \citep{cseh14}. Follow-up radio observations 1.5 years later showed that the core was resolvable with VLBI to a size of 16.4\,mas and had faded by more than a factor of 7,  likely caused by adiabatic expansion losses that dominate over synchrotron cooling. \citet{cseh15HoII} estimated that the core ejecta age is $\geq2.1$\,y but perhaps between $\sim13-46$ years. In contrast, the lobes remained at a more consistent flux density and are now brighter than the core, indicating they likely represent the longer-lived outer-most interaction of the jetted ejecta with the surrounding medium where their mechanical energy is converted to radiation \citep{cseh15HoII}. The discrete ejection events observed in Ho II X-1 suggest that ULX radio bubbles may be inflated by transient jets on timescales of only a few hundred years \citep{cseh14}. 
Given that \vtc{} and 43.78+59.3 have comparable radio luminosities to the combined core and lobe luminosity of Ho II X-1 as observed in 2012 \citep[$\nu L_{\nu} \approx 2 \times 10^{34}$\,erg\,s$^{-1}$;][]{cseh14}, 
it is possible these two transients could be younger versions of such a radio core and/or lobe nebula. 
Compact radio bubbles are also not unusual as one also has been observed around ULX IC 342 X-1. While this radio clump was unresolved with the VLA, it was resolved-out with VLBI observations indicating it is extended and likely nebular in nature \citep{cseh12}. 
  
\subsubsection{ULX radio bubble rates}
  
We now investigate whether the possibility of both \vtc{} and 43.78+59.3 being newly born ULX radio nebulae is consistent with the rates of known ULXs. If we limit the sample of ULXs to be within 5\,Mpc, the Updated Nearby Galaxy Catalog \citep{karachentsev13} lists 57 galaxies with reasonably high stellar masses of $>10^{8.49}$M$_{\odot}$ \citep[corresponding to the lowest mass galaxy with a known ULX in the][catalog]{swartz11} beyond the Local Group \citep[we used the $K$-band luminosity as the proxy for galaxy stellar mass; see][and references therein]{bell03}. Of these 57 galaxies, at least 14 contain ULXs, totaling at least 20 ULXs, including ULXs catalogued in \citet{swartz11}, the two ULXs in NGC 1313 \citep{colbert02}, and the three X-ray faint, extragalactic ULX radio bubbles mentioned above (NGC 7793-S26, M83 MQS1 and NGC 300-S10). This corresponds to $\gtrsim0.35$ ULXs per galaxy within 5\,Mpc. Of the 20 known extragalactic ULXs within 5\,Mpc, at least 6 have associated radio nebulae \citep[NGC 5408 X-1, NGC 7793-S26, IC 342 X-1, Ho II X-1, M83 MQS1 and NGC 300-S10;][]{kaaret03,soria06,pakull10,soria10,cseh12,miller05,soria14,urquart19},
corresponding to 30\% of the ULX population. This percentage is also consistent with the \citet{pakull08} estimate for extragalactic ULXs with shock-ionised optical bubbles, which are also tracers of strong ULX outflows (like jets or winds). This corresponds to $\gtrsim0.1$ ULXs with associated radio nebulae per galaxy within 5\,Mpc. 
If we assume that \vtc{} and 43.78+59.3 are indeed newly born ULX radio lobes, then by also including Ho II X-1, there are at least three known ULXs within 5\,Mpc that exhibit evidence of short-lived outflow events, corresponding to 0.05 per galaxy. This is only a factor of 2 lower than the number of persistent ULX radio nebulae within 5\,Mpc, which may suggest that some fraction of bubbles are inflated by the short-lived ejecta from sources such as \vtc{}, 43.78+59.3, and Ho II X-1.

\section{Conclusions}

The radio transient described in the paper is the first to be discovered as part of this VLA survey of M81, resulting in a preliminary surface density of $13.2^{+30.2}_{-10.9}$\,deg$^{-2}$ to a $3\sigma$ flux density limit of 75\,$\mu$Jy for this galaxy. However, the nature of the new radio transient is still unknown. With the available data, it is not possible to rule out a background active galaxy or long-term SN shockwave interactions with a dense circumstellar medium from an unusual SN similar to SN 1987A. 
The M81 transient \vtc{} is also more luminous than all but the very brightest radio flares observed from Galactic XRBs. However, its longevity and lack of short-term variability are at odds with such characteristics in the same Galactic systems and also the transient ULX in M31. The radio properties of \vtc{}, including the minimum energy, jet power, and magnetic field strength, 
are most closely matched to 43.78+59.3, discovered by \citet{muxlow10} in M82, likely associated with an accreting source, perhaps similar in nature to SS 433 and NGC 7793-S26 \citep{joseph11}. We go one step further to suggest that perhaps both \vtc{} and 43.78+59.3 are the birth of short-lived ULX radio lobes, similar to that observed near the core of Ho II X-1, implying a total of 0.05 such transients per galaxy within 5\,Mpc. Such an identification could potentially be a new class of extragalactic radio transient, where their ejecta may be partly responsible for the inflatation of some persistent ULX radio bubbles. 
In order to test this hypothesis, further radio follow-up is required to first determine if \vtc{} is still detectable and has an optically thin spectral index, demonstrating similar behaviour to 43.78+59.3. If \vtc{} were also still at the same flux density or (preferably) brighter, VLBI observations would determine if the source was also extended, as has been shown for 43.78+59.3 \citep{muxlow10}. Continued VLBI monitoring of both transients would allow us to determine if they have expanding radio nebulae similar to Ho II X-1, suggesting their identification as short-lived radio nebulae associated with super-Eddington accretion events.  

The detection of a radio transient through the M81 VLA monitoring campaign is a proof of concept for the ``targeted" observational technique and could therefore be an important strategy for SKA transient science.

\section*{Acknowledgements}

We thank the referee for their careful reading of the manuscript
and their recommendations.
GEA is the recipient of an Australian Research Council Discovery Early Career Researcher Award (project number DE180100346) and JCAM-J is the recipient of Australian Research Council Future Fellowship (project number FT140101082) funded by the Australian Government.
The scientific results reported in this article are based in part on observations made by the \cxo\ \textit{X-ray Observatory} as part of Chandra Award Number GO6-17078A (PI: D. Swartz).
RU acknowledges that this research is supported by an Australian Government Research Training Program (RTP) Scholarship.
TPR thanks the Science and Technology Facilities Council for funding as part of the consolidated grant ST/P000541/1.
The National Radio Astronomy Observatory is a facility of the National Science Foundation operated under cooperative agreement by Associated Universities, Inc.
This research has made use of data obtained from the Chandra Data Archive and the Chandra Source Catalog, and software provided by the Chandra X-ray Center (CXC) in the application package {\sc CIAO}.
Based on observations made with the NASA/ESA \textit{Hubble Space Telescope}, and obtained from the Hubble Legacy Archive, which is a collaboration between the Space Telescope Science Institute (STScI/NASA), the Space Telescope European Coordinating Facility (ST-ECF/ESA) and the Canadian Astronomy Data Centre (CADC/NRC/CSA).
This research has made use of NASA's Astrophysics Data System.
This research has made use of the SIMBAD database \citep{wenger00} and the VizieR catalogue access tool \citep{ochsenbein00} operated at CDS, Strasbourg, France. 
We acknowledge the use of NASA's SkyView facility \href{(http://skyview.gsfc.nasa.gov)}{http://skyview.gsfc.nasa.gov} located at NASA Goddard Space Flight Center \citep{mcglynn98}.
The following Python packages and modules were used for this research, including {\sc Astropy}, a community-developed core Python package for Astronomy \citep{TheAstropyCollaboration2013,TheAstropyCollaboration2018}, {\sc APLpy}, an open-source plotting package for Python \citep{robitaille_12_aplpy}, {\sc matplotlib} \citep{hunter07}, {\sc numpy} \citep{vanderWalt_numpy_2011} and {\sc scipy} \citep{Jones_scipy_2001}. 
This research has made use of SAOImage DS9, developed by Smithsonian Astrophysical Observatory.

\label{lastpage}

\bibliographystyle{mnras}
%\bibliography{bibliography}

\end{document}